\documentstyle[12pt]{article}
\newcommand{\newsection}{
\setcounter{equation}{0}
\section}
\def\appendix#1{
  \addtocounter{section}{1}
  \setcounter{equation}{0}
  \renewcommand{\thesection}{\Alph{section}}
  \section*{Appendix \thesection\protect\indent #1}
  \addcontentsline{toc}{section}{Appendix \thesection\ \ \ #1}
  }
\newcommand{\rf}[1]{(\ref{#1})}
\newcommand{\beq}{\begin{equation}}
\newcommand{\eeq}{\end{equation}}
\newcommand{\bea}{\begin{eqnarray}}
\newcommand{\eea}{\end{eqnarray}}

\newcommand{\la}{\left\langle}
\newcommand{\ra}{\right\rangle}

\newcommand{\eps}{\epsilon}

\newcommand{\calf}{{\cal F}}
\newcommand{\calg}{{\cal G}}

\newcommand{\calo}{{\cal O}}
\newcommand{\calp}{{\cal P}}
\newcommand{\cald}{{\cal D}}
\newcommand{\trace}{{\,\rm Tr \,}}

\begin{document}
\topmargin 0pt
\oddsidemargin 5mm
\headheight 0pt
\headsep 0pt
\topskip 9mm

\addtolength{\baselineskip}{0.20\baselineskip}

\begin{center}

\vspace{36pt}
{\large \bf
Mean Field Approach to the Giant Wormhole Problem
}

\vspace{36pt}

{\sl A. Gamba}
\footnote{
E-mail: gamba@vsmima.unimi.it.
}

\vspace{12pt}

Dipartimento di Matematica, Politecnico di Torino\\
Corso Duca degli Abruzzi 24, 10129 Torino, Italy\\

\vspace{24pt}

{\sl I. Kolokolov}
\footnote{
E-mail: kolokolov@inp.nsk.su.
}

\vspace{12pt}

Institute of Nuclear Physics \\
630090 Novosibirsk, Russia. \\

\vspace{24pt}

and

\vspace{24pt}

{\sl M.Martellini}
\footnote{
On leave of absence from:
Dipartimento di Fisica dell'Universit\`a di
Milano, Via Celoria 16, 20133 Milano, Italy and I.N.F.N., Sezione di Pavia,
27100 Pavia, Italy. \\
E-mail: martellini@milano.infn.it.
}

\vspace{12pt}

Dipartimento di Fisica, Universit\`a di Roma I {\it La Sapienza}\\
Piazzale Aldo Moro 2, 00185 Roma, Italy\\

\end{center}

\vspace{24pt}

\vfill

{\bf Abstract}

We introduce a gaussian probability density for the space-time distribution
of wormholes, thus taking effectively into account wormhole interaction.
Using a mean-field approximation for the free energy, we show that giant
wormholes are probabilistically suppressed in a homogenous isotropic
``large'' universe.

\vspace{12pt}

\noindent
\vspace{24pt}

\vfill

\newpage

\newsection{Introduction}

Some years ago it was observed
\cite{colgidstrom}
that topological fluctuations of space-time provide a mechanism which
can in principle fix the values of some fundamental constants of
nature.
More precisely, it appeared that
taking into account the possibility of creation of tiny
wormholes connecting distant regions of space-time
has the effect of modifying the coupling constants contained in
the model and to provide a probability distribution for them.
In particular, it was shown by Coleman that,
under certain assumptions, the probability distribution
for the cosmological constant
was infinitely peaked around zero, thus providing a candidate
solution, at least neglecting one-loop effects
\cite{polcarmart},
to the long-standing cosmological constant problem.

The main characteristic of wormholes is that the probability of
their creation does not depend on the positions of the endpoints
$x_0$
and
$x_1$,
at least when the two points are distant.
In a first approximation, one can think that the wormholes
are dilute, so that
$|x_1-x_0|$ is in the average great with respect
to the wormhole size
$a$,
and one can forget about the wormhole interaction
which is likely to appear when the endpoints become close to
each other.
This approximation was assumed in all previous analysis
of this subject.
However, it was evident from the very beginning
\cite{klebsussfish}
that this simplified model ran over some inconsistencies, like the
so-called giant wormhole paradox, that is, the non-physical prediction
that wormholes of macroscopic sizes should be generated.

In this paper we introduce a finer description of the process of wormhole
creation and destruction, and show that in this case giant wormholes are
probabilistically disfavoured.
More precisely, we assume that the
wormhole space-time distribution be described by the gaussian probability
density
\rf{three},
which corresponds to a suppression of the probability of creation of
a wormhole when the two endpoints come close to each other.
We then evaluate the functional integral for the theory using
methods which were previously introduced by one of us
\cite{kolokolov}
to deal with similar problems in solid state physics.
Using the mean-field approximation
\rf{twenty}
for the free energy, we show that the free energy for the theory is
proportional to
$\Lambda^4 a^2/\ell_P^2$
(where
$\Lambda$
is an energy scale
$\leq 1/\ell_P^2$
and
$a$
is the effective size of the wormholes), and therefore has a minimum
{\it for small values} of
$a$.
This is our main physical result, which implies that giant wormholes
are probabilistically suppressed in a homogenous isotropic ``large''
universe. This phenomenon is a large-scale, low-energy
``cooperative effect'' between the wormhole density
\rf{one}
and space-time
distribution
\rf{three}.
Previous analysis of the problem neglected a self-consistent
determination of the wormhole size
$a$
because they did not take
into account the wormhole
space-time distribution, arriving this way to the cited
giant wormhole paradox.
Furthermore, our analysis shows that
the result is quite independent from the
``microscopic details'' of the gravity-matter interaction, at
least in the domain of validity
($1/a<\Lambda$) of our model.

\newsection{Definition of the model}

As a starting point,
let us assume that we have performed integration over short-range
topology fluctuations.
It is natural to suppose that a reasonable trial wave function,
after eliminating such degrees of freedom, will give us some
effective theory of geometrical and matter degrees of freedom on the
wormholes background.
Moreover, it seems reasonable that the only trial parameter left
in the theory should be the effective size of the wormholes.
Thus, let us assume that the probability density for the creation of
a wormhole of size
$a$
be of the form
\beq
w_0=e^{-c a^2}
\label{one}
\eeq
where
$c={\rm const}=\calo (1)$
and the length
$a$
is measured in units of the Plank length
$\ell_P$.

Now, it is reasonable to think that the above
wormhole density distribution for finite values of
$a$
affects also the space-time distribution of the wormhole itself.

Of course, it is difficult to build a complete theory of the microscopic
dynamics of wormholes, but it seems reasonable as a first step to
describe this dynamics self-consistently by taking into account the size
$a$
both in the density and in the space-time correlations.
Let us in the following formulate a simple model of universe with
wormholes in which one may calculate explicitly the effect of the size
$a$, and the main result shall be that giant wormholes are probabilistically
suppressed.

For this purpose, we describe the space-time distribution
of the ``ends'' of the wormholes by the local field
\beq
\alpha(x)=\alpha\delta(x-x_0)
\label{two}
\eeq
and we postulate the following weight function for the field
$\alpha(x)$:
\beq
\calp [\alpha(x)]=
\exp (-{1\over 2}\int dx dx' \alpha(x)\alpha(x') K(x-x')),
\label{three}
\eeq
Here the correlation function
$K(x)$,
which is the amplitude for a wormhole insertion,
is assumed to be gaussian:
\beq
K(x)=K_0 \exp ( -a x^2) ,
\label{four}
\eeq
where
$K_0\sim\exp(-S_{\rm w})$
and
$S_{\rm w}$
is the (constant) wormhole action.
The {\it ansatz}
\rf{three}
allows us to take effectively into account
the finite size of wormholes in the space-time distribution.
As a matter of fact, if we substitute in
\rf{three}
the distribution
of two point-like wormholes
\beq
\alpha(x)=\alpha_0\delta(x-x_0)+\alpha_1\delta(x-x_1)
\label{five}
\eeq
we see that the distribution
\rf{three}
describes a
suppression of probability when
$|x_1-x_0|<a$,
and we get factorization when
$|x_0-x_1|\rightarrow\infty.$

As a first order of approximation one may think that the field
$\alpha$,
which describes the large-scale structure of the space-time
distribution of wormholes, is coupled to gravity by volume effects
only.
In other words, we may assume that the microscopic structure of the
gravitational field is not relevant to describe the
size distribution of the
wormhole
dynamics.
Therefore the gravitational degrees of freedom associated with the
spin-2 part, i.e.
$h_{\mu\nu}=g_{\mu\nu}-\la g_{\mu\nu}\ra$,
which control the small-scale structure, can be neglected at this
level.
As a consequence of this picture, we impose the following coupling of
$\alpha(x)$ with gravity:
\beq
S_{\rm int}=\int d^4 x\sqrt{g(x)}\, \alpha(x),
\label{nine}
\eeq
where the ``volume field''
$\sqrt{g}$
is expanded as
\beq
\sqrt{g(x)}=1+\varphi(x), \qquad |\varphi|\ll 1.
\label{ten}
\eeq
The model that we have in mind is given by the path integral
\bea
&&
\int\cald\alpha(x)\,
\exp\left(
-{1\over 2}
\int d^4x\, d^4x'\, \alpha(x) K(x-x') \alpha(x')
\;+\; S_{\rm int}
\right)  \nonumber \\
&\cong&
{1\over\sqrt{\det K}}
\exp\left(
{1\over 2}
\int d^4x\, d^4x'\,\varphi(x) K^{-1}(x-x')\varphi(x')
\right)
\label{pathintegral}
\eea
where
\rf{pathintegral}
is defined up to a normalization and we have excluded the
term linear in
$\varphi$.
Indeed this contribution may be neglected if we integrate
\rf{pathintegral}
over
$\varphi$,
since it represents an arbitrary fluctuation field around the vacuum
$\la\sqrt{g}\ra=1$.
Of course, looking in this way, we are taking into account quantum
gravity volume effects with a functional measure given by
$\exp(-S(\varphi))$,
where
$S(\varphi)$ is the effective action
\beq
S(\varphi)={1\over 2}\int dxdx'\,\varphi(x)K^{-1}(x-x')\varphi(x'),
\label{eleven}
\eeq
coming from the above functional integration over
$\alpha$.
We get Coleman's picture when we consider
only the global degrees of freedom of the field $\alpha(x)$.
For instance, if we substitute into the
$\cald\alpha$-integral the factor
\beq
\delta(\alpha-\int d^4x\,\alpha(x))
\label{fourteen}
\eeq
and perform the
$\cald\alpha$-integration,
we would have a theory with the global variable
$\alpha$ playing the role of an effective
wormhole collective coordinate in Coleman's spirit.

In any case, the effective model based on
\rf{eleven}
is only ultralocal for
$\varphi$.
In a realistic scenario of quantum gravity coupled with matter fields we
must investigate also the effect of the density and space-time wormhole
distribution over the coupled system matter-gravity.
So, a less crude model is obtained if we add to the non-local action
\rf{eleven}
a matter field contribution
$S_{\rm mat}$
represented by a massless Klein-Gordon field coupled to the geometry
$g_{\mu\nu}$.
Assuming again that only volume effects are relevant, we may replace
the covariant Laplacian in the
$S_{\rm mat}$
with the flat one, and using
\rf{ten},
$S_{\rm mat}$ looks as
\beq
S_{\rm mat}=\int d^4x\,\rho(x)(1+\varphi(x))\Delta\,\rho(x)
\label{fifteen}
\eeq
where
$\Delta$
is the flat, Euclidean 4-dimensional Laplacian.

The coupled model is now described by
$S(\varphi)+S_{\rm mat}(\rho,\varphi)$.
Performing firstly the
$\rho$-functional
integration, we get a free energy of the form (in Euclidean signature)
\beq
F[\varphi]\simeq{1\over 2}\log\det[(1+\varphi(x))\,\Delta ].
\label{starstar}
\eeq
Of course we must understand the functional determinant in
\rf{starstar}
as a renormalized one. For this aim in the following section we
shall adopt a Schwinger-DeWitt proper-time scheme, together
with an ``$1\over D$ expansion''.

\newsection{Mean Field Approximation}

The complete quantum partition function of our model is given by
\bea
Z &=&
\int\cald\varphi\,
\exp\left(
-{1\over 2}
\int d^4x\, d^4x'\, \varphi(x) K^{-1}(x-x')\varphi(x')-F[\varphi]
\right) \nonumber	\\
&=&
\la \exp\left({-F[\varphi]}\right)\ra_\varphi
\label{treuno}
\eea
where the average symbol means
\beq
\la\calo\ra_\varphi\equiv\int\cald\varphi\,\calo\,
\exp\left(
-{1\over 2}\int\varphi K^{-1}\varphi
\right)
\label{tredue}
\eeq

At this point we use the mean-field approximation
\beq
Z\simeq \exp\left({-\la F[\varphi]\ra_\varphi}\right)
\label{tretre}
\eeq
Let us remind that, due to the convexity of the exponential
function, one has the Peierls inequality
\beq
\la e^A\ra\geq e^{\la A\ra}.
\eeq
Here we assume to take the lower bound of
$\la e^A\ra$,
since in our case
$A$
is given by a functional
$-F[\varphi]$
of
$\varphi$,
and the kernel
\rf{four},
goes very rapidly to zero for
$|x-y|\rightarrow\infty$;
this is just the large-range approximation
underlying all our model.

In other words, the true free energy
$\calf\equiv -\log(Z)$
is approximated by
\beq
\calf\simeq
\langle F[\varphi]{\rangle}_\varphi
= \la{1\over 2}\log\det [(1+\varphi)\Delta]\ra_\varphi.
\label{trequattro}
\eeq
By using now the Schwinger-De Witt proper-time formalism, we may write
\bea
&& \log\det[(1+\varphi)\,\Delta]	\nonumber \\
&=& -\int^\infty_\epsilon {dt\over t} \trace \exp
\left(
-t[(1+\varphi)\,\Delta]
\right),\qquad\epsilon\rightarrow 0^+
\label{trecinque}
\eea
where $\trace$ stands for the space time trace of the heat-kernel
operator in
\rf{trecinque}
and can be represented as a Schr\"odinger-like path integral over closed
paths according to
\bea
&&\trace \exp\left(
-t[(1+\varphi)\,\Delta ]
\right) \\
&&\qquad=\int d^4x\,\int_{x(0)=x(t)=x}\cald x(\tau)\,
\exp\left(
-\int_0^t d\tau {1\over 4} (1+\varphi[x(\tau)])\,\dot x(\tau)^2
\right).	\nonumber
\label{tresei}
\eea
Putting
\rf{trecinque}
and
(3.7)
in
\rf{trequattro},
we get
\bea
\calf
&\simeq& \lim_{\epsilon\rightarrow 0^+}
	\int\cald\varphi
	\exp
	\left(
		-{1\over 2}\int dxdx'\varphi(x)K(x-x')\varphi(x')
	\right)
\,\cdot\,	\\
&&\cdot\,
\int d^4x\,\int_\eps^\infty{dt\over t}\int_{x(0)=x(t)=x}\,\cald x(\tau)
\exp
	\left\{
		-\int_0^t
		\left(
			{1\over 4}(1+\varphi[x])
			\dot x(\tau)^2
		\right)
	\right\},	\nonumber
\label{tresette}
\eea
At this point, the functional integration can be done
\cite{kolokolov},
giving the expression
\bea
\calf
&\simeq&
\lim_{\epsilon\rightarrow 0^+}
-{1\over 2}\int d^4x\, \int_\epsilon^\infty\,
{dt\over t} \int_{x(0)=x(t)=x} \cald x(\tau)\,\cdot	\\
&&
\cdot\,
\exp\left(
-\int_0^t d\tau\,{\dot x(\tau)^2\over 4} +
{1\over 2}\int_0^t\int_0^t d\tau d\tau'\,
\dot x(\tau)^2  K[x(\tau)-x(\tau')] \dot x(\tau')^2
\right)	\nonumber
\label{eighteen}
\eea
It is important that the path integral over
$\cald x(\tau)$
is translationally invariant and that the integration over
$d^4 x$
gives simply the volume factor
$V$
(actually a covariant volume with respect to a background metric
$\hat g_{\mu\nu}$
such that
$\sqrt{\hat g}\equiv 1$):
\bea
\calf
&\simeq&
-{1\over 2} V f, \nonumber \\
f
&=&
\lim_{\epsilon\rightarrow 0^+}
\int_\eps^{+\infty}{dt\over t}\int_{x(0)=x(\tau)=0}\cald x(\tau)\,\cdot  \\
&&\cdot
	\exp\left\{-\int_0^t d\tau {\dot x^2\over 4}+
	{1\over 4}\int_0^t\int_0^td\tau d\tau'
		\dot x(\tau)^2 K(x(\tau)-x(\tau') \dot x(\tau')^2)\right\}.
\nonumber
\label{eighteenshtrikh}
\eea

Path integrals of this kind can be found in the theory of disordered systems.
Unfortunately, it is impossible to calculate exactly even this very
simplified path integral. However, we can estimate it in some sense
non-perturbatively and obtain rough numerical estimates and an enough
good qualitative picture.

Really, our
$x(\tau)$
lies in 4-dimensional space, here assumed isotropic and homogenous,
and it seems to be reasonable that
the main contribution to the path integral
\rf{eighteenshtrikh}
is given by trajectories having high degree of isotropy.
For such trajectories:
\bea
<x(\tau)x(\tau')>	&=&	0,	\nonumber	\\
<(x(\tau)x(\tau'))^2>	&=&	{1\over D} <x^2(\tau)>^2
\label{nineteen}
\eea
where
$D$
is the dimension of the space.
In our case
$D=4$,
and we can formulate the mean-field approximation for the path
integral
\rf{eighteenshtrikh}
by requiring that:
\bea
K(x(\tau)-x(\tau')) &=& c a^2\exp\left(-{1\over a^2}
		(x(\tau)-x(\tau'))^2\right)		\nonumber	\\
	&\approx& c a^2\exp\left(-{1\over a^2} x^2(\tau)\right)
			\exp\left(-{1\over a^2} x^2(\tau')\right).
\label{twenty}
\eea
(It is important to notice that this decomposition becomes exact in the limit
$a\rightarrow\infty$;
see below).
In this approximation the calculation of
\rf{eighteenshtrikh}
can be reduced to solving the quantum mechanical problem
\bea
f &\simeq&
\lim_{\epsilon\rightarrow 0^+}
	 {1\over \sqrt{\pi c a^2}}
	\int_\eps^{+\infty}{dt\over t}
	\int_{-\infty}^{+\infty}ds e^{-s^2/c a^2}\,\cdot  	\nonumber\\
&&\qquad\qquad\cdot
	\int_{x(0)=x(\tau)=0}\cald x(\tau)
		\exp\left\{-\int_0^t d\tau\left[
		{\dot x^2\over 4}+ s \dot x^2 e^{-x^2/a^2}\right]\right\}
					\nonumber	\\
  &=& \lim_{\epsilon\rightarrow 0^+}
	\int_\eps^{+\infty}{dt\over t}
		\int_{-\infty}^{+\infty} ds e^{-s^2/c a^2}
			G_s(t|0,0),
\label{twentyone}
\eea
where
$G_s(t|x,x')$
is the Green function
\bea
\partial_t G_s(t|x,x') &=&
	{1\over 4}(1+4 s e^{-x^2/a^2})
	\,\Delta G_s(t|x,x'),		\\
G_s(0,x,x') &=& \delta^{(4)}(x-x').
\nonumber
\label{twentytwo}
\eea

Some remarks are in order. In our
mean-field
approximation
the fluctuations of gravity
are described by the {\it global} variable
$s$,
which has the meaning of a sort of order parameter.
Indeed the
parameter
$s$
can be considered as an amplitude of the volume-field
$\sqrt{g}$
fluctuations.
It seems that the very existence of an order parameter of this kind has
more general meaning and more deep origin than our approximation.
We see from
\rf{twentytwo}
that at
$s=-1/4$
we will have
$G_s(t|x,x')$
very singular at
$x\rightarrow x'$.
Such fluctuations can be described as bags in space-time, where
$\sqrt{g}\rightarrow 0$.

On the other hand, the domain
$s<-1/4$
corresponds to fluctuations with
$\sqrt{g}<0$,
i.e., we have
a first-order phase transition, since also the covariant volume factor
$V\equiv\int d^4x\,\sqrt{\hat g}$
in
\rf{eighteenshtrikh}
changes sign, and hence we pass from a regime, say, with
$\calf>0$
to one entropy dominated, i.e.
$\calf<0$.
{}From a quantum-mechanical point of view, a negative free energy should
imply vacuum decay, so that in order to avoid this
possibility we restrict ourselves to the domain of integration
$s>-1/4$
(i.e.,
$\sqrt{g}>0$).

We come thus to the final formulation of our model: the volume density
of free energy has the form:
\bea
f
&&\simeq\lim_{\epsilon\rightarrow 0^+}
\int_\eps^{+\infty}{dt\over t}\int_{-1/4}^{+\infty}ds
	e^{-s^2/c a^2}\cdot\sqrt{2\over\pi c a^2}\,\cdot \\
&&\qquad\qquad
	\cdot\int_{x(0)=x(\tau)=0}\cald x(\tau)
	\exp\left\{-\int_0^t\left({\dot x^2\over 4}+s \dot x^2 e^{-x^2/a^2}
				\right)\right\}
\nonumber
\label{twentyoneshtrikh}
\eea
In
\rf{twentyoneshtrikh}
is understood the usual zero-point energy subtraction of
${\rm Tr\;log\;}\Delta\equiv f_0$.

\newsection{General Estimates}

A limiting case in which
\rf{twentyoneshtrikh}
can be computed is obtained by setting
$a\rightarrow +\infty$.
In this limit we can neglect the
$x$-dependence of the correlator, i.e. we can put:
\beq
e^{-x^2/a^2}\simeq 1
\label{twentythree}
\eeq
Then, one gets
\beq
f\simeq
\lim_{\epsilon\rightarrow 0^+}
	\int_\eps^{+\infty}{dt\over t} \sqrt{2\over \pi c a^2}
	\int_{-1/4}^{+\infty}ds\, e^{-s^2/c a^2}
	\cdot <0|e^{-t(1+4s)\Delta}|0> - f_0
\label{twentyfour}
\eeq
This quantity requires an ultraviolet regularization. Let us
use a momentum cut-off
regularization by defining
\beq
<0|e^{-t(1+4s)\Delta}|0>=
{2\pi^2\over(2\pi)^4}\int_0^\Lambda k^3 dk\, e^{-t k^2(1+4s)},
\label{twentyfourshtrikh}
\eeq
where
$\Lambda$
is the UV-cutoff.
By
\rf{twentyfourshtrikh},
the leading contribution to
\rf{twentyfour}
in
$\Lambda$
is approximately equal to (see Appendix):
\beq
f\approx{1\over \pi}\Lambda^4 c a^2 8(1-{2\over 3}\sqrt{2\over\pi})
	\approx{4\over\pi^2}\Lambda^4 c a^2.
\label{twentyfive}
\eeq
Since in our model we have neglected systematically all quantum gravity
short-range effects, we may assume that the energy-scale
$\Lambda$
is upper-bounded by the mass Planck scale
$1/\ell_P$
(in
$c=\hbar=1$).

\newsection{The General Case}

For arbitrary
$a$
the quantum mechanical problem
\rf{twentyoneshtrikh}
cannot be solved
exactly, but we see from the previous computation that the main
contribution is given by high-momentum fluctuations and that we can
use semiclassical formulas.
It is convenient to use Laplace (or Fourier) transform.

The Laplace transform of the Green function
\rf{twentytwo}
in the
$t$-variable
has a cut along the imaginary semiaxis; the inverse Laplace
transform can be reduced to an integral of the jump of
$G$
on this cut over this semiaxis. The resulting expression, after
$dt$-integration,
has the form:
\beq
f\simeq
	\int_0^\infty{d\omega\over\pi}\log\omega
	\int_{-1/4}^{+\infty} ds e^{-s^2/c a^2}\sqrt{2\over \pi c a^2}
	\cdot{\rm Im}\, G_s(\omega+i0|0,0)-f_0,
\label{twentyseven}
\eeq
where
$G_s(\omega+i0|x,x')$
obeys the equation
\bea
&&(V(x)\Delta+\omega) G_s(\omega|x,x')=\delta^{(4)}(x-x'); \\
&& V(x)\equiv 1+\varphi(x)
\nonumber
\label{twentyeight}
\eea
The Green function
$G_s(\omega|x,0)$
is spherically symmetric:
\beq
G_s(\omega|x,0)=\calg_s(\omega,r),
\label{twentynine}
\eeq
and taking into account the small imaginary part of
$\omega$
we obtain its semiclassical limit in the form:
\beq
\calg_s(\omega,r)\simeq-1{k(0)\over 8\pi V_0 r}
	H^{(1)}_1\left(\int_0^rdr'\, k(r')\right),
\label{thirty}
\eeq
Here we have set
\beq
k(r)=\sqrt{2\omega\over 1+4s e^{-r^2/a^2}},\qquad
V_0\equiv 1+4s
\label{thirtyone}
\eeq
and
$H^{(1)}_1$
is the Hankel function of first kind.
It is easy to see that the value of
${\rm Im}\,\calg_s(\omega,r\rightarrow 0)$
is the same as in the case
$a\rightarrow\infty$
and that we obtain the same result
\rf{twentyfive}.
As a consequence the
main contribution would be given by the domain
$k(0)<\Lambda$,
i.e.
$\tilde s t>1/\Lambda$
in the notations of the Appendix.
It should be stressed that the mean field approximation becomes
exact in the limit
$a\rightarrow\infty$.
On the other hand,
$a$
is large for the dominating fluctuations.
Thus, we can hope that the main contribution to the free energy in the
giant wormhole limit is taken into account by
\rf{twentyfive}.

The cutoff
$\Lambda$
has physical meaning and is defined by quantum gravity effects.
As we have noticed above
it is natural to think that
$\Lambda\sim1/\ell_P$.
Our estimations have relation to the case that
$1/a<\Lambda<1/\ell_P$
only.
In this limit we see that the volume density of free energy increases with
$a$
as
$\Lambda^4 a^2$,
{\it and the energy minimum is realized at small}
$a$.

\bigskip
\centerline{\bf Acknowledgments}

One of us (I.~K.) would like to acknowledge the Department of Physics
of the University of Milan, where part of this work was done,
for warm hospitality.
\bigskip

\newsection{Appendix}

We estimate the integral
\rf{twentyfour},
\rf{twentyfourshtrikh}.
Let
$ s= {1\over 4}(\tilde s-1)$,
then:
\bea
f\simeq{1\over 4\pi^2}\sqrt{2\over\pi c a^2}
&&	\int_\eps^{+\infty}{dt\over t}
	\int_0^\infty d\tilde s\,\exp\left(-{1\over 16 c a^2}(\tilde s-1)^2
								\right)
						\nonumber 	\\
&&
	\cdot{1\over\tilde s^2 t^2}
		\left\{\left(1-e^{-{t\tilde s\over 4}\Lambda^2}\right)-
		{t\tilde s\over 4}\Lambda^2 e^{-{t\tilde s\over 4}\Lambda^2}
						\right\}
	-f_0.
\eea
We divide the domain of
$dt\, d\tilde s$-integration in two domains,
$\tilde s t<1/\Lambda^2$
and
$\tilde s t>1/\Lambda^2$.
The first domain gives a contribution which cancels the
$f_0$, while the second gives
\bea
f &\simeq& {1\over 4\pi^2}\sqrt{2\over\pi c a^2}
	\int_\eps^{+\infty}{dt\over t}
	\int_{1/\Lambda^2 t}^{+\infty} d\tilde s \,
	\exp\left(-{1\over 16 c a^2}(\tilde s-1)^2\right)
	\cdot{1\over \tilde s^2 t^2}	\nonumber	\\
&\simeq&
	\int_{1/(4\Lambda^2\sqrt{c a^2})}
		^{1/\Lambda^2} {dt\over t^3}
	 	\left(1-{1\over 4\Lambda^2 t}\sqrt{2\over\pi c a^2}\right)
\simeq{4\over\pi^2 \Lambda^4 c a^2},\qquad a^2\gg 1.
\eea
The domains of
$dt$-integration
$1/\Lambda^2<t$
and
$t<1/(4\Lambda^2\sqrt{c a^2})$
give a small contribution.




\newpage

\begin{thebibliography}{99}
\bibitem{colgidstrom}
S.~Coleman, {\it Nucl. Phys.} {\bf B307} (1988) 867; \\
S.~Giddings and A.~Strominger, {\it Nucl. Phys.} {\bf B307} (1988) 854.
\bibitem{polcarmart}
J.~Polchinski, {\it Phys. Lett. B} {\bf 219} 251; \\
A.~Carlini and M.~Martellini, {\it Phys. Lett. B} {\bf 276} 36.
\bibitem{klebsussfish}
I.~Klebanov, L.~Susskind and T.~Banks, {\it Nucl. Phis.} {\bf B317} 665; \\
W.~Fishler and L.~Susskind, {\it Phys. Lett. B} {\bf 217} (89) 48.
\bibitem{kolokolov}
I.~Kolokolov, {\it Phys. Lett.} {\bf 123} (87) 302.
\end{thebibliography}
\end{document}